# Compatibility of Maxwell's fluid equations with interactions between oscillating bubbles

Ion Simaciu[a,*], Gheorghe Dumitrescu[b], Zoltan Borsos[c,**] and Viorel Drafta[d]

[a] Retired lecturer, Petroleum-Gas University of Ploiești, Ploiești 100680, Romania

[b] Retired professor, High School Toma N. Socolescu, Ploiești, Romania

[c] Petroleum-Gas University of Ploiești, Ploiești 100680, Romania

[d] Independent researcher

E-mail: * isimaciu@yahoo.com; ** borzolh@upg-ploiesti.ro

Abstract

The result of this paper is the derivation of the expression for the interaction force between two pulsating/oscillating bubbles (the secondary Bjerknes force). This is done by expressing the force using the Maxwell hydrodynamic equations for a fluid. These subsequent equations were written for the quantities velocity, pressure, and density as deviations from steady state. Also, as it will show in the following rows, we found the expressions of the acoustic charge and of the acoustic intensity. Our results may facilitate an approach for the interaction of two vortices and for the interaction of a bubble with a vortex. This approach leads to a model for an acoustic charged particle which has internal angular momentum.

## 1. Introduction

The analogy between the Maxwell equations and those of the compressible fluid has a long time ago observed [1-8]. Following this analogy, we infer the secondary Bjerknes force [9-12] and set the background of "Acoustic world" Project AWP [13-18]. Our project tries to argue that the world of the systems which interact, and correlate trough acoustic waves is like the world where systems interact and correlate trough electromagnetic waves. This similarity take place theoretically and experimentally.

We assume the acoustic world to be a fluid whose temperature is close to the boiling temperature, and which is found into a container. Here may rise some specific systems like acoustic waves, bubbles, and vortices or we can introduce solid corpuscles and drops from another liquid. They are the basic systems - phenomena which can initiate more complex systems like standing acoustic waves, bubble trapped in a vortex (bubble with spin or acoustic particle with spin), systems of two acoustic particles (acoustic "atoms") and systems with a large number of bubbles – clusters of bubbles. We must mention that we took care to set our models to be invariant to the change of the different coordinate frames. Since we address our study to those systems which interact and corelate acoustically then we will adopt the limit of the velocity of the waves to be the velocity of the acoustic waves which propagate into an unperturbed fluid ($u = (dp/d\rho)_s$). The choice mentioned above is depicted by Lorentz transformations. We note that we do not refer here to the relativistic theory of the hydrodynamics [19, Ch. 15]]. Equations of hydrodynamic are invariant to the transformations of Galilei but equations of electromagnetism are invariant to the Lorentz transformations [2].

This is a new stage of the basics of the theory of the AWP and which consist in depiction of the correlations between fluid physics, acoustic theory, the theory of bubbles and vortices [20-25] and of the topic addressed to the study of the analogy between mechanics and electromagnetism.. More precisely, first we will prove that it is possible to infer secondary Bjerknes force [9-11] force from the hydrodynamics' equations of the type of Maxwell. Earlier we have proven [12, 17, 18] approaches which depict the analogy between the interactions bubble-bubble and of those of charged particles without spin [31, 32]. Now this analogy will allow us to model a charged acoustic particle with angular momentum gained by entrapment of the bubble by a vortex [23-25].

When the viscosity of the fluid is large the small vortices (minivortices) have short lifetime $\tau_l$ ($\tau_l \ll R_c/u$, $R_c$ is radius of the container where acoustic phenomenon take place) and therefore the interactions between them and other minisystems found in container are difficult to be observed. The task becomes easier if in container is found a superfluid (a fluid cooled so much as the viscosity becomes extremely small) and the container is found into a space station. In this case vortices are stable and quantified [21, 22]. The large lifetimes of the quantified vortices allow to make more complex structures. The missing of gravity in space facilitates experimental observations of the phenomena instead of the terrestrial labs where observations are influenced by gravity.

Theoretical and experimental study of acoustic phenomena will allow a better understanding of the electromagnetic phenomena and vice versa [2]. Also, a better understanding will be for quantification, for how elementary unpunctual particles can be generated, for the role played by the limit of the magnitude of the velocity which is involved in interactions and depends on the properties of the fluid, for how laws which govern phenomena can be relative and also for observation.

In the second section we will depict the shape of secondary Bjerknes force and of the acoustic intensity using the analogy with the Maxwell equations. There we will outline what is the acoustic charge. We will use the SI units and Gaussian units. Also, there we write the Maxwell equations for a compressible fluid. In the third section we infer the expressions of the acoustic charge, of the acoustic intensity and of the acoustic force, for all of them using Maxwell equations for fluids. The fourth section is addressed to conclusions.

## 2. Maxwell's equations and the Bjerkenes secondary force

### 2.1. Clarifications on Maxwell's equations for the electrostatic field

In the Electromagnetic World, i.e. the world where interactions propagate at the maximum velocity of light in a vacuum, $c = 1/\sqrt{\varepsilon_0 \mu_0}$., Maxwell's equations for the electrostatic field in a vacuum are, in the SI units [26]

$$\nabla \vec{E} = \frac{\rho_q}{\varepsilon_0} \quad (1)$$

where the electric charge density, $\rho_q = dq/dV$, is the source of the electrostatic field $\vec{E}$ and $\varepsilon_0$ is the permittivity of the vacuum. We assume that vacuum is the medium where electromagnetic interactions propagate. There, the electric field acts upon a charged particle with a force

$$\vec{F} = q\vec{E} \qquad (2)$$

or the force with which a charged particle acts to another charged particle is

$$\vec{F}_{ij} = \frac{q_i q_j \hat{r}}{4\pi\varepsilon_0 r^2} = q_i \left( \frac{q_j \hat{r}}{4\pi\varepsilon_0 r^2} \right) = q_i \vec{E}_j, \hat{r} = \frac{\vec{r}}{r} \qquad (3)$$

The above equations are asymmetric since the charge does not depend on the properties of the medium and the intensity depends on the properties of the medium where interactions propagate.

We can define symmetric equations by introducing a charge which depends on properties of the medium:

$$\vec{F}_{sij} = \frac{e_i e_j \hat{r}}{r^2} = e_i \left( \frac{e_j \hat{r}}{r^2} \right) = e_i \vec{E}_{sj}, \vec{E}_{sj} = \frac{e_j \hat{r}}{r^2}. \qquad (4)$$

$$\vec{F}_s = e\vec{E}_s, \vec{E}_s = \frac{e\hat{r}}{r^2} \qquad (5)$$

$$\nabla \vec{E}_s = 4\pi \rho_s, \rho_s = \frac{de}{dV} \qquad (6)$$

These two shapes of the forces lead to a relation between the parameters involved here since forces are equal, $\vec{F}_s = \vec{F}, e\vec{E}_s = q\vec{E}$. Therefore, one writes:

$$e^2 = \frac{q^2}{4\pi\varepsilon_0}, e = \frac{q}{\sqrt{4\pi\varepsilon_0}}, \vec{E}_n = \vec{E} = \frac{q\hat{r}}{4\pi\varepsilon_0 r^2} = \frac{e\hat{r}}{\sqrt{4\pi\varepsilon_0} r^2} = \frac{\vec{E}_s}{\sqrt{4\pi\varepsilon_0}} \qquad (7)$$

The above equations are Maxwell equations in the Gaussian units [26].

## 2.2. Secondary Bjerknes forces

Secondary Bjerknes forces which act between two bubbles which oscillate are [11, 12]:

$$\vec{F}_B = \langle \vec{F}_{ij} \rangle = -\langle V_i(t)(\nabla p_j(r,t)) \rangle; \ i \neq j = 1,2; \qquad (8)$$

with: $V_i(t) = 4\pi R_i^3(t)/3$ and with the pressure exerted by the acoustic waves radiated by bubbles $p_j(r,t) = \rho_0 \ddot{V}_j/(4\pi r) - (\rho_0 \dot{R}_j^2/2)(R_j/r)^4$, $\ddot{V}_j = \partial_t^2 V_j = \partial^2 V_j/\partial t$. We used the symbol $\langle .. \rangle$ which designates the operation to average of a parameter during a period. In the approach which we adopted the variable volume of the bubble means the acoustic charge and the gradient of the pressure means the intensity of the acoustic field around the bubble. The pressure is the scalar potential of the acoustic field. Acoustic charge is a measure of the property of the bubble to scatter and to absorb some of the energy of the inductor wave [10, 12, 18]. Bubble absorbs some of the energy of the inductor wave and radiates spherical acoustic waves of the same frequency as of the inductor wave. This phenomenon is a scattering. The absorption is due to the viscosity of the liquid and of the viscosity of the inner gas of the bubble, i.e., thermal viscosity.

When $r >> R_i$ one can approximate the pressure to be

$$p_j(r,t) = \rho_0 \left[ \frac{\ddot{V}}{4\pi r} - \frac{\dot{R}^2}{2}\left(\frac{R}{r}\right)^4 \right] \cong \rho_0 \frac{\ddot{V}_j}{4\pi r}. \tag{9}$$

Then, the acoustic force may be expressed as

$$\vec{F}_B = \langle \vec{F}_{ij} \rangle_t = \left\langle V_i \left( \rho_0 \frac{\ddot{V}_j \hat{r}}{4\pi r^2} \right) \right\rangle_t = \frac{\rho_0 \hat{r}}{4\pi r^2} \langle V_i \ddot{V}_j \rangle_t \tag{10}$$

We note that the Eq. (10) is asymmetric regarding to the way of how the temporal variable volumes acts in the expressions of acoustic charge, $V_i(t)$, and of the acoustic intensity $\rho_0 \ddot{V}_j(t)\hat{r}/(4\pi r^2)$.

For oscillations with small amplitude [11, 12] one can express the temporal variability of the radius of the bubble as:

$$R_i(t) = R_{0i}\left[1 + a_i \cos(\omega t + \varphi_i)\right], \ddot{R}_i(t) = -\omega^2 a_i R_{0i} \cos(\omega t + \varphi_i), a_i \ll 1 \tag{11}$$

and $\ddot{V}_i = 4\pi\left(2\dot{R}_i^2 + R_i^2 \ddot{R}_i\right) \cong 4\pi R_i^2 \ddot{R}_i = (4\pi R_{0i}^3/3)\left[-3\omega^2 a_i \cos(\omega t + \varphi_i)\right] = V_{0i}\left[-3\omega^2 a_i \cos(\omega t + \varphi_i)\right]$, since $2\dot{R}_i^2 \ll R_i \ddot{R}_i$.

With this approximation, the pressure expression becomes

$$p_j(r,t) \cong \frac{\rho_0 \ddot{V}_j}{4\pi r} \cong \frac{\rho_0 R_j^2 \ddot{R}_j}{r} \cong \frac{V_{0j}}{4\pi r}\left[-3\rho_0 \omega^2 a_j \cos(\omega t + \varphi_j)\right] \tag{12}$$

and the volume as

$$V_i = \frac{4\pi R_i^3}{3} = \frac{4\pi R_{0i}^3 \left(1 + a_i \cos(\omega t + \varphi_i)\right)^3}{3} \cong V_{0i}\left(1 + 3a_i \cos(\omega t + \varphi_i)\right). \tag{13}$$

To derive the expression of the acoustic force one substitutes Eq. (13) in Eq. (8) and so one obtains

$$\vec{F}_B = \langle \vec{F}_{ij} \rangle = \left\langle V_{0i}\left(1 + 3a_i \cos(\omega t + \varphi_j)\right)\left[\frac{V_{0j}\left(-3\rho_0 \omega^2 a_j \cos(\omega t + \varphi_j)\right)\hat{r}}{4\pi r^2}\right]\right\rangle \cong$$
$$-\left\langle V_{0i}\left(3a_i \cos(\omega t + \varphi_j)\right)\left[\frac{V_{0j}\left(3a_j \cos(\omega t + \varphi_j)\right)\hat{r}}{4\pi \left(\frac{1}{\rho_0 \omega^2}\right) r^2}\right]\right\rangle, \tag{14}$$

since $\left\langle V_{0i}V_{0j}\left[\left(-3\rho\omega^2 a_j \cos(\omega t + \varphi_j)\right)\hat{r}/(4\pi r^2)\right]\right\rangle = 0$. Here, in Eq. (14) we identify $q_{ai} = \rho_0 \delta V_i = \rho_0 (V_i - V_{0i})$ to be the acoustic charge of the bubble, which is the variation in time of the volume of the bubble. This variation also involves the variation of the virtual mass or induced mass, $\delta m_{bi} = \rho_0 \delta V_i /2 = 3\rho_0 V_{0i} a_i \cos(\omega t + \varphi_j)/2$, $m_{bi} = \rho_0 V_{0i}/2$, [11, 20]

$$q_{ai}(r,t) = \rho_0 \delta V_i = 2\delta m_i = 3V_{0i}\rho_0 a_i \cos(\omega t + \varphi_j). \tag{15}$$

Acoustic charge is interpreted as the property of the pulsating bubble to scatter acoustic waves. Also, we identify the intensity of the acoustic field around the bubble which oscillates. It is:

$$\vec{E}_{qaj}(r,t) = \frac{-3V_{0j}\rho_0 a_j \cos(\omega t + \varphi_j)\hat{r}}{4\pi(\rho_0/\omega^2)r^2} = \frac{-\rho_0 \delta V_j \hat{r}}{4\pi(\rho_0/\omega^2)r^2} = \frac{-q_{aj}\hat{r}}{4\pi\varepsilon_a r^2}, \quad (16)$$

and it depends on the acoustic charge $q_{aj} = \rho_0 \delta V_j$ and on the acoustic permittivity

$$\varepsilon_a = \frac{\rho_0}{\omega^2}. \quad (17)$$

Acoustic permittivity depends on the properties of acoustic medium which is excited by an acoustic wave, i.e. the liquid which has the $\rho_0$ density in the stationary state and which oscillates with pulsation $\omega$ under the action of the inductor wave.

Following the above definitions, the acoustic force may be expressed as

$$\vec{F}_{Bij} = \left\langle \frac{-\rho_0 \delta V_i \delta V_j \hat{r}}{4\pi\varepsilon_a r^2} \right\rangle = \left\langle \frac{-q_{ai}q_{aj}\hat{r}}{4\pi\varepsilon_a r^2} \right\rangle = \left\langle q_{ai}\vec{E}_{qaj} \right\rangle. \quad (18)$$

The same acoustic force may be expressed by the product of two acoustic charges which depends on the properties of the fluid. To do this we identify in Eq. (14) an acoustic charge which depends on the fluid which oscillates

$$e_{ai} = \frac{q_{ai}}{\sqrt{4\pi\varepsilon_a}} = \frac{\omega q_{ai}}{\sqrt{4\pi\rho_0}} \quad (19)$$

and an intensity of the acoustic field which has the shape

$$\vec{E}_{eaj} = \frac{-e_{aj}\hat{r}}{r^2}. \quad (20)$$

With the above definitions the acoustic force becomes

$$\vec{F}_{Bij} = \left\langle \frac{-e_{ai}e_{aj}\hat{r}}{r^2} \right\rangle_t = \left\langle e_{ai}\vec{E}_{eaj} \right\rangle_t. \quad (21)$$

There are papers [27] where the expressions of the acoustic force have a symmetric shape regarding the dependence of the acoustic charge and the intensity of the field on the time

$$\vec{F}_{Bs} = \left\langle \vec{F}_{sij} \right\rangle = -\left\langle \left(\dot{V}_i(t)\right)\left(\frac{\rho_0 \dot{V}_j(t)\hat{r}}{4\pi r^2}\right) \right\rangle_t = \frac{-\rho_0 \hat{r}}{4\pi r^2}\left\langle \dot{V}_i\dot{V}_j \right\rangle_t. \quad (22)$$

In this case the acoustic charge is $\dot{V}_i(t)$ and the intensity of the acoustic field generated by the oscillating bubble is $\rho_0 \dot{V}_j(t)\hat{r}/(4\pi r^2)$. When adopting this shape of the acoustic force one loose the hydromechanics meaning of the way to generate the force by the action of a bubble which press another bubble, as we find in (8). But if we perform a temporal average we are led to the same expression of the force because $\left\langle \left(\dot{V}_i(t)\right)\left(\dot{V}_j(t)\right)\right\rangle_t = \omega^2(V_{0i}3a_i)(V_{0j}3a_j)\cdot \left\langle \sin(\omega t+\varphi_i)\sin(\omega t+\varphi_j)\right\rangle_t = -\left\langle (V_i(t))(\ddot{V}_j(t))\right\rangle_t$ up to the terms which are proportional with $a_i a_j$.

### 2.3. Analogous Maxwell's equations for the compressible fluid case

According to the paper [1], the Fluid Maxwell equations for compressible fluid with vortex are:

$$\nabla \cdot \vec{H}_a = 0, \quad (23)$$

$$\nabla \cdot \vec{E}_a = 4\pi \rho_a, \tag{24}$$

$$\nabla \times \vec{E}_a + \frac{\partial \vec{H}_a}{\partial t} = 0, \tag{25}$$

$$u^2 \left(\nabla \times \vec{H}_a\right) - \partial_t \vec{E}_a = \vec{J}_a \equiv \frac{4\pi}{u} \vec{j}_a, \tag{26}$$

where the vectors of the field have the following definitions

$$\vec{E}_a \equiv -\frac{\partial \vec{\upsilon}}{\partial t} - \nabla h, \ \vec{H}_a \equiv \vec{\omega} = \nabla \times \vec{\upsilon}. \tag{27}$$

In Eqs. (23-27): $u$ is the velocity of sound in the fluid in a uniform state at rest, $\rho_a$ the acoustic charge density and $\vec{J}_a = (4\pi/u)\vec{j}_a = (4\pi/u)\rho_a \vec{\upsilon}$ the acoustic current density, defined according to the acoustic field variables $\vec{\upsilon}$ and $h = p/\rho_0$. The charge density and the current density vector are like the scalar potential $\phi$ and to the vector potential $\vec{A}$ from electrodynamics. These later parameters may be defined as:

$$4\pi \rho_a = -\frac{\partial (\nabla \vec{\upsilon})}{\partial t} - \Delta h, \tag{28}$$

$$\vec{J}_a \equiv \frac{4\pi}{u} \vec{j}_a = -\frac{\partial^2 \vec{\upsilon}}{\partial t^2} + \nabla\left(\frac{\partial h}{\partial t}\right) + u^2 \left(\nabla \times (\nabla \times \vec{\upsilon})\right). \tag{29}$$

These equations are correct for phenomena in the fluid when the flow velocity is much lower than the mechanical wave propagation velocity in that fluid and this is much lower than the electromagnetic wave velocity in the fluid: $\upsilon \ll u \ll c_f$. For high flow rates, $\upsilon \leq u \leq c_f$, pressures and temperatures for which the fluid is relativistic, the equations of relativistic fluid dynamics are valid [19, Ch. 15]. In the papers Gauge Symmetries in Physical Fields (Review) and Fluid Gauge Theory, Kambe derives Maxwell's equations for the relativistic fluid [33]. According to Fluid Gauge Theory, the Maxwell equations that describe the interactions between specific liquid systems (for example water, at pressure $p = 10^5 \, \text{Nm}^{-2}$, is liquid when the temperature is in the interval $T \in [273-373] \, \text{K}$) and phenomena (the motions of the centers of mass of specific systems: vortices, bubbles, drops, corpuscles, etc.) must be invariant only with respect to the wave velocity in the liquid $u$ and so in the relativistic Maxwell equations for the liquid we make the substitutions: $c_f \to u, \beta = \upsilon/c_f \to \beta_f = \upsilon/u$ ($\upsilon$ is the flow velocity or the velocity of the mass centers of the specific systems).

In the paper Analogy between vortex waves and EM waves [2], Jamati decomposes the quantities that characterize the fluid, $\vec{\upsilon}, p, \rho$, into quantities at the equilibrium state, $\vec{\upsilon}_0, p_0, \rho_0$, and quantities that represent the deviation from the equilibrium state, $\vec{\upsilon}_d, p_d, \rho_d$, according to the relations:

$$\vec{\upsilon} = \vec{\upsilon}_0 + \vec{\upsilon}_d, \ p = p_0 + p_d, \ \rho = \rho_0 + \rho_d. \tag{30}$$

Using the decomposition one obtains equations which are similar to Maxwell equations for those variables which correspond to deviation from equilibrium, $\vec{\upsilon}_d, p_d, \rho_d$:

$$\nabla \cdot \vec{E}_d = (\nabla \cdot \vec{\alpha}_d) = \frac{\rho_{ad}}{\varepsilon_d}, \tag{31}$$

$$\nabla \cdot \vec{B}_d = \left(\nabla \cdot \vec{\beta}_d\right) = 0, \tag{32}$$

$$\nabla \times \vec{E}_d + \frac{\partial \vec{B}_d}{\partial t} = \left(\nabla \times \vec{\alpha}_d + \frac{\partial \vec{\beta}_d}{\partial t}\right) = 0, \tag{33}$$

$$\left(\nabla \times \vec{B}_d\right) - \frac{\partial \vec{E}_d}{u^2 \partial t} = \left(\left(\nabla \times \vec{\beta}_d\right) - \frac{\partial \vec{\alpha}_d}{u^2 \partial t}\right) = \vec{J}_d \equiv \mu_d \vec{j}_d. \tag{34}$$

where the vectors of the field, $\left[\vec{E}_d\right]_{SI} = [v_d]_{SI} \cdot \left[\vec{B}_d\right]_{SI}$, are defined as:

$$\vec{E}_d = \vec{\alpha}_d \equiv -\frac{\partial \vec{v}_d}{\partial t} - \frac{\nabla p_d}{\rho_0}, \quad \vec{B}_d = \vec{\beta}_d \equiv \nabla \times \vec{v}_d. \tag{35}$$

In Eqs. (31-35): $u$ is the velocity of sound in the fluid in a uniform state at rest, $\rho_{ad}$ is the acoustic charge density and $\vec{j}_d \equiv \rho_{ad} \vec{v}_d$ is the acoustic current density (a current vector in [1]) for the deviation from the equilibrium state:

$$\frac{\rho_{ad}}{\varepsilon_d} = \frac{\partial^2 (p_d)}{\rho_0 u^2 \partial t^2} - \frac{\nabla^2 p_d}{\rho_0} + \frac{\partial (\vec{v}_0 \cdot \nabla \rho_d)}{\rho_0 \partial t}, \tag{36}$$

$$\vec{J}_{ad} \equiv \mu_d \vec{j}_d = \frac{\partial^2 \vec{v}_d}{u^2 \partial t^2} - \nabla^2 \vec{v}_d - \frac{\nabla (\vec{v}_0 \cdot \nabla \rho_d)}{\rho_0} \tag{37}$$

and they are defined by the help of field variables $\vec{v}_d$, $p_d = \rho_0 u v_d$, the density $\rho_d = \rho_0 v_d / u$ [19 - Sch. 63] and the constants: acoustic permittivity $\varepsilon_d$ and acoustic permeability $\mu_d$, $1/\sqrt{\varepsilon_d \mu_d} = u$.

### 2.4. Analogous Maxwell's equations in the vortex-free fluid

If the vortex is missing in fluid, $\vec{v}_0 = \vec{\Omega} \times \vec{r} = 0$, $\vec{B}_d = \vec{\beta}_d \equiv \nabla \times \vec{v}_d = 0$, then the vector $\vec{v}_d$ has only a radial component and therefore the Maxwell equations for fluid (31-35) become:

$$\nabla \cdot \vec{E}_d = \left(\nabla \cdot \vec{\alpha}_d\right) = \frac{\rho_{ad}}{\varepsilon_d}, \tag{38}$$

$$\nabla \times \vec{E}_d = \left(\nabla \times \vec{\alpha}_d\right) = 0, \tag{39}$$

$$-\frac{\partial \vec{E}_d}{u \partial t} = \left(-\frac{\partial \vec{\alpha}_d}{u \partial t}\right) = \mu_d \vec{j}_d. \tag{40}$$

Consequently, the vector $\vec{E}_d$ (35) and the acoustic charge density $\rho_{ad}$ (36) are:

$$\vec{E}_d = \vec{\alpha}_d \equiv -\frac{\partial \vec{v}_d}{\partial t} - \frac{\nabla p_d}{\rho_0} = -\frac{\hat{r}}{\rho_0 u} \frac{\partial p_d}{\partial t} - \frac{\nabla p_d}{\rho_0}, \quad \left[\vec{E}_d\right]_{SI} = \text{ms}^{-2}; \tag{41}$$

$$\frac{\rho_{ad}}{\varepsilon_d} = \left[\frac{\partial^2 p_d}{\rho_0 u^2 \partial t^2} - \frac{\nabla^2 p_d}{\rho_0}\right], \quad \left[\frac{\rho_{ad}}{\varepsilon_d}\right]_{SI} = \text{s}^{-2}. \tag{42}$$

It is noted that both the acoustic field intensity, Eq. (41), and the charge density, Eq. (42), contain a term that depends on the explicit time pressure variation.

## 3. Inferring secondary Bjerknes force from analogous Maxwell equations

According to the analogous Maxwell equations when vortex is missing in fluid the acoustic force is:

$$\vec{F}_a = q_a \vec{E}_a \qquad (43)$$

with

$$q_a = \int \rho_a 4\pi r^2 dr, \quad [q_a]_{SI} = \text{kg}. \qquad (44)$$

One can find out the shape of the expression of the secondary Bjerknes force, tacking account to the hydrodynamic Maxwell equations, if first we infer the expressions of the acoustic charge of a bubble $q_{bi} = \int \rho_{abi} 4\pi r^2 dr$ and of acoustic vector intensity $\vec{E}_{bj}$. To do this, we assume that in a fluid which has the density $\rho_0$ at equilibrium there is a bubble with radius $R_0$ which oscillates due to the pressure of a plane wave with angular velocity $\omega$. This motion of the bubble generates around it an acoustic field having the pressure $p_b$, according to Eq. (9)

$$p_b(r,t) = \rho_0 u \upsilon_b = \rho_0 \left[ \frac{\ddot{V}}{4\pi r} - \frac{\dot{R}^2}{2}\left(\frac{R}{r}\right)^4 \right] \cong \frac{\rho_0 \ddot{V}}{4\pi r}. \qquad (45)$$

When substitute the pressure from Eq. (45) in Eq. (41) it results the acoustic intensity of the field being around the bubble

$$\vec{E}_{bj} = -\frac{\hat{r}}{\rho_0 u}\frac{\partial p_{bj}}{\partial t} - \frac{\nabla p_{bj}}{\rho_0} = -\frac{\ddot{V}_j \hat{r}}{4\pi u r} + \frac{\ddot{V}_j \hat{r}}{4\pi r^2}. \qquad (46)$$

Similar, when substitute Eq. (45) in Eq. (42), it results the acoustic charge density around the bubble which oscillates

$$\frac{\rho_{bi}}{\varepsilon_d} = \frac{\partial^2 (\ddot{V}_i)}{4\pi r u^2 \partial t^2} - \frac{\nabla^2}{4\pi}\left(\frac{\ddot{V}_i}{r}\right) = \frac{\ddddot{V}_i}{4\pi r u^2} + \rho_{0bi}. \qquad (47)$$

In Eq. (47) $\rho_{0bi}$ is the acoustic charge density of the bubble which oscillates, for $r \leq R_i$. It is introduced to be like the case of the electric charge distributed into a sphere with finite radius $R$. This constant $\rho_{0bi}$ was introduced because $(\ddot{V}_i/4\pi)\nabla^2(1/r) = (\ddot{V}_i/4\pi)[(1/r^2)(\partial/\partial r)(r^2((\partial/\partial r)(1/r)))] = (-\ddot{V}_i/4\pi)[(1/r^2)(\partial(1)/\partial r)] = -0$ and then the charge density is equal to the dimensional constant $\rho_{0bi}$ ($[\rho_{0bi}]_{SI} = \text{s}^{-2}$) for $r \leq R_i$. For $r > R_i$ the charge density is $\ddddot{V}_i/(4\pi r u^2)$.

According to Eq. (44), the charge of the oscillating bubble is:

$$q_{bi} = \int_0^r \rho_{bi} 4\pi r^2 dr = \varepsilon_d \int_0^r \left(\frac{\ddddot{V}_i}{4\pi r u^2} + \rho_{0bi}\right) 4\pi r^2 dr =$$

$$\varepsilon_d \int_0^{R_i} \rho_{0bi} 4\pi r^2 dr + \varepsilon_d \int_{R_i}^r \left(\frac{\ddddot{V}_i}{4\pi r u^2}\right) 4\pi r^2 dr = V_i \varepsilon_d \rho_{0bi} - \frac{\ddddot{V}_i \varepsilon_d R_i^2}{2u^2} + \frac{\ddddot{V}_i \varepsilon_d r^2}{2u^2}. \qquad (48)$$

We set the limit of the integration to be $r > R_i$ because the absorption and the scatter due to the inductor wave occurs in the inner gas bubble of radius $R_i$, according to the section 2.2. Substituting Eq. (46) and (48) in Eq. (43) one can find out the shape of temporal vector force:

$$\vec{F}_{bij}(r,t) = q_{bi}\vec{E}_{bj} = \left(\varepsilon_d V_i \rho_{0bi} - \frac{\varepsilon_d \ddddot{V}_i R_i^2}{2u^2} + \frac{\varepsilon_d \ddddot{V}_i r^2}{2u^2}\right)\left(\frac{\ddot{V}_j \hat{r}}{4\pi r^2} - \frac{\dddot{V}_j \hat{r}}{4\pi u r}\right) =$$

$$\frac{\varepsilon_d \rho_{0bi} V_i \ddot{V}_j \hat{r}}{4\pi r^2} - \frac{\varepsilon_d \rho_{0bi} V_i \dddot{V}_j \hat{r}}{4\pi u r} - \frac{\varepsilon_d \ddddot{V}_i R_i^2 \ddot{V}_j \hat{r}}{8\pi u^2 r^2} + \frac{\varepsilon_d \ddddot{V}_i R_i^2 \dddot{V}_j \hat{r}}{8\pi u^3 r} + \tag{49}$$

$$\frac{\varepsilon_d \ddddot{V}_i \ddot{V}_j \hat{r}}{8\pi u^2} - \frac{\varepsilon_d \ddddot{V}_i \dddot{V}_j r \hat{r}}{8\pi u^3}.$$

When average the temporal force during a period $T = 2\pi/\omega$, the force becomes independent of time:

$$\vec{F}_{bij}(r) = \left\langle \vec{F}_{bij}(r,t)\right\rangle = \left\langle \frac{\varepsilon_d \rho_{0bi} V_i \ddot{V}_j \hat{r}}{4\pi r^2}\right\rangle - \left\langle \frac{\varepsilon_d \rho_{0bi} V_i \dddot{V}_j \hat{r}}{4\pi u r}\right\rangle -$$

$$\left\langle \frac{\varepsilon_d \ddddot{V}_i R_i^2 \ddot{V}_j \hat{r}}{8\pi u^2 r^2}\right\rangle + \left\langle \frac{\varepsilon_d \ddddot{V}_i R_i^2 \dddot{V}_j \hat{r}}{8\pi u^3 r}\right\rangle + \left\langle \frac{\varepsilon_d \ddddot{V}_i \ddot{V}_j \hat{r}}{8\pi u^2}\right\rangle - \left\langle \frac{\varepsilon_d \ddddot{V}_i \dddot{V}_j r \hat{r}}{8\pi u^3}\right\rangle \cong \tag{50}$$

$$\frac{\varepsilon_d \rho_{0bi} \hat{r}}{4\pi r^2}\left\langle V_i \ddot{V}_j\right\rangle - \frac{\varepsilon_d \hat{r}}{8\pi u^2 r^2}\left\langle \ddddot{V}_i R_i^2 \ddot{V}_j\right\rangle + \frac{\varepsilon_d \hat{r}}{8\pi u^2}\left\langle \ddddot{V}_i \ddot{V}_j\right\rangle,$$

because $\left\langle V_i \dddot{V}_j\right\rangle \cong 0$, $\ddddot{V}_i R_i^2 \dddot{V}_j \cong 0$ and $\left\langle \ddddot{V}_i \dddot{V}_j\right\rangle \cong 0$ up to the terms which are proportional with $a_i a_j \ll 1$. This force has physical dimensions $\left[F_{bij}\right]_{SI} = \text{kgms}^{-2}$.

When compare Eq. (50) with Eq. (10) we can assume that the first term in Eq. (50) to be Bjerknes force if:

$$\varepsilon_d = \frac{\rho_0}{\rho_{0bi}} = \frac{\rho_0}{\omega^2}, \left[\rho_{0bi}\right]_{SI} = \text{s}^{-2}, \rho_{0bi} = \omega^2, \tag{51}$$

i.e. the permittivity $\varepsilon_d$ is the acoustic permittivity (17). Substituting Eqs. (51) in Eq. (50) we obtain the expression of the force between two pulsating bubbles (bubbles in an oscillating fluid!), according to Maxwell's hydrodynamic equations

$$\vec{F}_{bij}(r) \cong \frac{\rho_0 \hat{r}}{4\pi r^2}\left\langle V_i \ddot{V}_j\right\rangle - \frac{\rho_0 \hat{r}}{8\pi u^2 \omega^2 r^2}\left\langle \ddddot{V}_i R_i^2 \ddot{V}_j\right\rangle + \frac{\rho_0 \hat{r}}{8\pi u^2 \omega^2}\left\langle \ddddot{V}_i \ddot{V}_j\right\rangle. \tag{52}$$

Finally, if we substitute in Eq. (52) the secondary Bjerknes force (10), it results:

$$\vec{F}_{bij}(r) \cong \vec{F}_{Bij}(r) - \frac{\rho_0 \hat{r}}{8\pi u^2 \omega^2 r^2}\left\langle \ddddot{V}_i R_i^2 \ddot{V}_j\right\rangle + \frac{\rho_0 \hat{r}}{8\pi u^2 \omega^2}\left\langle \ddddot{V}_i \ddot{V}_j\right\rangle. \tag{53}$$

This force which acts between two oscillate bubbles comprise a first term which is the Bjerknes force, the second term which represents a force which depends inversely proportional to the square of the distance between bubbles and proportional to the square of the radius of the $i$ bubble $R_i^2$ and the third term which does not depend on the distance between the two bubbles. The second term of the force is asymmetric $\vec{F}_{b2ij}(r) \neq \vec{F}_{b2ji}(r) = \left(\rho_0 \left\langle \ddddot{V}_i R_j^2 \ddot{V}_j\right\rangle \hat{r}\right)/\left(8\pi u^2 \omega^2 r^2\right)$ regarding the dependence of the radii of those two bubbles and break the law of action and reaction. The third term of the force is symmetric, $\vec{F}_{b3ij}(r) = \vec{F}_{b3ji}(r) = \left[\rho_0 \hat{r}/\left(8\pi u^2 \omega^2\right)\right]\left\langle \ddddot{V}_j \ddot{V}_i\right\rangle$. These force terms was not yet depicted theoretically and highlight experimentally when one study the interactions between oscillate bubbles. As we showed in this paper the Maxwell's hydrodynamic equations, for parameters which are addressed to the deviations from the equilibrium, $\vec{v}_d, p_d, \rho_d$, have led to the existence of an asymmetric force and a force independent of distance. This fact maybe requires a deeper study regarding the way of how the approximations were made when inferring these equations.

Using the same procedure as in section 2.2, one can express those two forces as function of acoustic charge defined by Eq. (13) and Eq. (19). With the definitions mentioned above the factors $\langle V_i \ddot{V}_j \rangle_t$, $\langle \dddot{V}_i R_i^2 \ddot{V}_j \rangle_t$ and $\langle \dddot{V}_i \ddot{V}_j \rangle$ can be approximated as:

$$\langle V_i \ddot{V}_j \rangle_t \cong -\omega^2 \langle \delta V_i \delta V_j \rangle = -\omega^2 \langle q_{ai} q_{aj} \rangle, \langle \dddot{V}_i R_i^2 \ddot{V}_j \rangle_t \cong -\omega^6 R_{0i}^2 \langle \delta V_i \delta V_j \rangle = \\ -\omega^6 R_{0i}^2 \langle q_{ai} q_{aj} \rangle, \langle \dddot{V}_i \ddot{V}_j \rangle \cong -\omega^6 \langle \delta V_i \delta V_j \rangle = -\omega^6 \langle q_{ai} q_{aj} \rangle. \tag{54}$$

With these expressions, the force (52) becomes:

$$\vec{F}_{bij}(r) = \frac{-\omega^2 \langle q_{ai} q_{aj} \rangle \hat{r}}{4\pi \rho_0 r^2} + \frac{R_{0i}^2 \omega^4 \langle q_{ai} q_{aj} \rangle \hat{r}}{8\pi \rho_0 u^2 r^2} - \frac{\omega^4 \langle q_{ai} q_{aj} \rangle \hat{r}}{8\pi \rho_0 u^2} = \\ \frac{-\omega^2 \langle q_{ai} q_{aj} \rangle \hat{r}}{4\pi \rho_0 r^2} \left(1 - \frac{R_{0i}^2 \omega^2}{2u^2}\right) - \frac{\omega^4 \langle q_{ai} q_{aj} \rangle \hat{r}}{8\pi \rho_0 u^2} \tag{55}$$

or with Eq. (19), one obtains

$$\vec{F}_{bij}(r) = \frac{-\langle e_{ai} e_{aj} \rangle \hat{r}}{r^2}\left(1 - \frac{R_{0i}^2 \omega^2}{2u^2}\right) - \frac{\omega^2 \langle e_{ai} e_{aj} \rangle \hat{r}}{2u^2}. \tag{56}$$

From the expression of the natural angular velocity, $\omega_{0i}^2 = p_{eff}/(\rho_0 R_{0i}^2)$, [11, 12], result $R_{0i}^2 = p_{eff}/(\omega_{0i}^2 \rho_0)$ and $R_{0i}^2 \omega^2/(2u^2) = (\omega^2/\omega_{oi}^2)(p_{eff}/2\rho_0 u^2)$. It follows that for the case $\omega \leq \omega_{0i}$ the term $R_{0i}^2 \omega^2/(2u^2) \ll 1$ and therefore the asymmetric component of the force of interaction of the two bubbles is much smaller than the secondary Bjerknes force. The term independent of the distance between the two bubbles represents a force constant equal to the secondary Bjrkenes force at a distance proportional to the wavelength of the inductive oscillating field $r = \sqrt{2}(u/\omega) = \lambda/(\sqrt{2}\pi)$.

## 4. Conclusions

As we expected to take place, the Maxwell equations for fluids can be used to infer the force of interactions between oscillate bubbles. Similarly, to the electric forces inferred from Maxwell equations one can infer the expression of the secondary Bjerknes force from Maxwell equations for compressible fluids. What is different is that we derived an expression which is asymmetric regarding the dependence of the radii of the bubbles $\vec{F}_{bij}(r) \neq \vec{F}_{bji}(r) = \left(\rho_0 \langle \dddot{V}_j R_j^2 \ddot{V}_i \rangle \hat{r}\right)/(8\pi u^2 \omega^2 r^2)$. There is also a force that does not depend on the distance between the two bubbles. These additional terms probably appear as a result of the approximations performed to infer Maxwell equations for parameters which measures the deviations from the equilibrium $\vec{v}_d, p_d, \rho_d$ and from the fact that bubble is an approach of acoustic charge without intern angular momentum.

Both in classic theory [28] and quantum mechanics [29] the electron is a particle without intern angular moment. We consider that only the study of a bubble trapped by a vortex can be a more complete model of an acoustic charge with angular moment. The approach of the electron as a vortex [30] can argue for our assumption.

The compatibility of the Maxwell equations for fluids with the properties of the interactions of systems specific to a fluid, i.e., oscillating/ pulsating bubbles and vortices, can be strong support for the assumption of the theoretic and experimental analogy between acoustic world and electromagnetic world.

## References

[1] Tsutomu Kambe, A new formulation of equations of compressible fluids by analogy with Maxwell's equations, Fluid Dyn. Res. 42, 055502, 2010.
[2] Fady Jamati, Analogy between vortex waves and EM waves, Fluid Dyn. Res. 50 065511, 2018.
[3] Arbab, A. I., The analogy between electromagnetism and hydrodynamics, Physics Essays 24, 254-259, 2011.
[4] Elena A. Ivanova, Modeling of electrodynamic processes by means of mechanical analogies. Z Angew Math Mech. 101:e202000076, 2021.
[5] Valery P. Dmitriev, Mechanics of electromagnetic interactions, arXiv:physics/0612210.
[6] Valery P. Dmitriev, Towards an Exact Mechanical Analogy of Particles and Fields, Nuov.Cim. 111 A, N5, pp.501-511, 1998, https://doi.org/10.1007/BF03185584.
[7] Wang, X. S., Derivation of Maxwell's equations based on a continuum mechanical model of vacuum and a singularity model of electric charges. Prog. Phys. 2, 111–120, 2008.
[8] Lachezar S. Simeonov, Mechanical Model of Maxwell's Equations and of Lorentz Transformations Foundations of Physics, volume 52, article number 52, 2023.
[9] Bjerknes, V. F. K. 1906, Fields of Force. Columbia University Press.
[10] Doinikov, Alexander A., Acoustic radiation forces: Classical theory and recent advances, Recent Res. Devel. Acoustics, 1, 39-67, 2003.
[11] T. Barbat, N. Ashgriz, and C. S. Liu, Dynamics of two interacting bubbles in an acoustic field, J. Fluid Mech. 389, 137-168, 1999.
[12] I. Simaciu, Z. Borsos, Gh. Dumitrescu, G. T. Silva and T. Bărbat, The acoustic force of electrostatic type, Bul. Inst. Politeh. Iaşi, Secţ. Mat., Mec. teor., Fiz. 65 (69), No 2, pp. 15-28, 2019; arXiv:1711.03567v1, 2017.
[13] I. Simaciu, Z. Borsos, Gh. Dumitrescu, A. Baciu, Planck-Einstein-de Broglie type relations for acoustic waves, arXiv:1610.05611, 2016.
[14] I. Simaciu, Z. Borsos, A. Baciu and G. Nan, The Acoustic World: Mechanical Inertia of Waves, Bul. Inst. Politeh. Iaşi, Secţ. Mat., Mec. teor., Fiz. 62 (66), No 4, pp. 52-63, 2016.
[15] I. Simaciu, Gh. Dumitrescu, Z. Borsos and M. Brădac, Interactions in an Acoustic World: Dumb Hole, Adv. High Energy Phys.,Vol. 2018, article ID 7265362, 2018.
[16] I. Simaciu, Gh. Dumitrescu, Z. Borsos, Acoustic lens associated with a radial oscillating bubble, Bul. Inst. Politeh. Iaşi, Secţ. Mat., Mec. teor., Fiz. 66 (70), No 2, pp. 9-15, 2020; arXiv:1811.08738, 2018.

[17] I. Simaciu, Gh. Dumitrescu and Z. Borsos, Mach's Principle in the Acoustic World, arXiv: 1907.05713, 2019; Buletinul Institutului Politehnic din Iași, Secția Matematică. Mecanică Teoretică. Fizică, Volumul 67 (71), No. 4, 59-69, 2021.
[18] Ion Simaciu, Zoltan Borsos, Viorel Drafta and Gheorghe Dumitrescu, Phenomena in bubbles cluster, arXiv:2212.12790, 2023.
[19] L. D. Landau, E. M. Lifshitz, Fluid Mechanics, Vol. 6, Third Rev. Ed., 1966.
[20] Tsutomu Kambe, Elementary Fluid Mechanics, EFM-Nankai - 2005.
[21] M. R. Matthews, B. P. Anderson, P. C. Haljan, D. S. Hall, C. E. Wieman and E. A. Cornell, Vortices in a Bose-Einstein Condensate, Phys. Rev. Lett. 83 (13), pp. 2498 – 2501, 1999.
[22] Weiler, C. N., Neely, T. W., Scherer, D. R., Bradley, A. S., Davis, M. J. and Anderson, B. P., Spontaneous vortices in the formation of Bose-Einstein condensates, Nature 455 (7215), 948–951, 2009.
[23] L.Y. Chena, L.X. Zhanga and X.M. Shaoa, The motion of small bubble in the ideal vortex flow, Procedia Engineering 126, pp. 228 – 231, 2015.
[24] Oweis, G. F.; van der Hout, I. E.; Iyer, C.; Tryggvason, G.; Ceccio, S. L., Capture and inception of bubbles near line vortices, Physics of Fluids 17(2): 022105-022105-14. http://hdl.handle.net/2027.42/87832, 2005.
[25] Victor P. Ruban, Bubbles with attached quantum vortices in trapped binary Bose-Einstein condensates, Journal of Experimental and Theoretical Physics, Vol. 133, No. 6, pp. 779-785, https://doi.org/10.1134/S1063776121120062, 2021.
[26] Jackson J. D., Classical Electrodynamics, 2nd ed., Wiley, New York, 1975.
[27] Rasoul Sadighi-Bonabi, Nastaran Rezaee, Homa Ebrahimi and Mona Mirheydari, Interaction of two oscillating sonoluminescence bubbles in sulfuric acid, Physical Review E 82, 016316, 2010.
[28] Barut, A. O. and Zanghi, N., Classical Model of the Dirac Electron, Phys. Rev. Lett. 52, pp. 2009-2012, 1984.
[29] Paul A. M. Dirac, The quantum theory of the electron, Proc. Roy. Soc. Lond. A, 117, 610-624, 1928.
[30] S. C. Tiwari, Anomalous magnetic moment and vortex structure of the electron, Modern Physics Letters A, Vol. 33, No. 31, 1850180, 2018.
[31] Zavtrak, S. T., A classical treatment of the long-range radiative interaction of small particles, Journal of Physics A, General Physics 23 (9), 1493, 1999.
[32] A. A. Doinikov and S. T. Zavtrak, Radiation forces between two bubbles in a compressible liquid, J. Acoust. Soc. Am. 102 (3), 1997.
[33] Tsutomu Kambe, Global Journal of Science Frontier Research: A Physics and Space Science, Volume 21, Issue 4, Version 1, 2021.